# Next4: Snapshots in Ext4 File System

Aditya Dani[1], Shardul Mangade[2], Piyush Nimbalkar[3], Harshad Shirwadkar[4]

*Computer Engineering, Pune Institute of Computer Technology,
Pune-411043, Maharashtra, India*

[1]`aditya.dani@gmail.com`, [2]`shardul.mangade@gmail.com`, [3]`piyushmnimbalkar@gmail.com`,

[4]`harshadshirwadkar@gmail.com`

*Abstract-* **The growing value of data as a strategic asset has given rise to the necessity of implementing reliable backup and recovery solutions in the most efficient and cost-effective manner. The data backup methods available today on Linux are not effective enough, because while running, most of them block I/Os to guarantee data integrity. We propose and implement Next4 - file system based snapshot feature in Ext4 which creates an instant image of the file system, to provide incremental versions of data, enabling reliable backup and data recovery. In our design, the snapshot feature is implemented by efficiently infusing the copy-on-write strategy in the write-in-place, extent based Ext4 file system, without affecting its basic structure. Each snapshot is an incremental backup of the data within the system. What distinguishes Next4 is the way that the data is backed up, improving both space utilization as well as performance.**

## 1. INTRODUCTION

### 1.1. Motivation:

In any storage system, the feature that is of paramount importance to the user is data protection. While the user takes for granted that data will not be lost due to a system crash or a design bug, the user also expects that a storage system will protect him or her from inadvertent deletions, unwanted modifications, malicious agents, etc.

Various solutions have been proposed by the storage industry, and have been adapted to various degrees. One very common method is by taking regular backups, either on the storage system itself, or onto tapes. This method has the major disadvantage of requiring massive storage capacity, and additionally, of causing substantial disruption in service while a backup is being made. Additionally, because of these disadvantages, it is not possible to create backups that are more frequent than about once a day, causing a large window of loss if data must be restored.

### 1.2. Snapshot Concept:

A snapshot is a point-in-time copy of a volume similar to taking a picture of the data at that instant. Snapshots look and behave like complete backups – they can be mounted as volumes, read simultaneously without affecting the volume that they are snapshots of, rolled back onto the volume if necessary, and deleted to free space. The ease and flexibility with which these operations may be performed has been instrumental in speeding up the adoption of snapshot technology in the IT industry.

Snapshots can be implemented at volume level as well by a volume manager such as LVM/LVM2[1]. Logical Volume Manager (LVM)/LVM2 is a volume manager located between a file system and a device driver. The snapshot in LVM is achieved by block- level copy-on-write (COW) and is performed on a unit of Logical Volume in the Volume Group. However, the snapshot capabilities at the LVM level have some caveats as below:

- The need to define a fixed volume size per snapshot.
- Copy overhead per snapshot for every write operation.
- Not scalable to O(1TB), requires O(1GB) RAM and long loading time.
- There is some memory overhead per mounted snapshot volume.
- Not possible to define a snapshot on a directory (sub volume).

Some of these caveats can be tackled by changes to the LVM snapshot implementation, but some are fundamental to the volume level snapshot approach. For this reason it makes sense to try to reach the design goals by implementing snapshots at the file system level.

### 1.3. Next4 Snapshots:

Next4 deals with implementation of file system based snapshot feature in Ext4 file system, without disturbing the basic Ext4 structure and features. Ext4 is the successor of the popular and most widely used Ext3 file system. Next4 will enable to create an incremental, read only snapshot of the file system, which can be recovered back by mounting the snapshot file on a loop back device. Next4 snapshots provide the simplicity of the LVM snapshots with high performance and scalability. The snapshot management capabilities include creating as many snapshots as desired and deleting any past snapshots to free up file system space.

## 2. DESIGN DETAILS

**Snapshot File**

Every snapshot taken is stored in a snapshot file, which is a regular sparse file. A sparse file is a type of computer file that attempts to use file system space more efficiently when blocks allocated to the file are mostly empty. This is achieved by writing brief information (metadata) *representing* the empty blocks to disk instead of the actual 'empty' space which makes up the block, using less disk space. A snapshot file has size same as that of the entire file system and represents the state of the file system when the snapshot was taken. Every logical block offset in the snapshot file represents a physical block in the underlying block device. A mapped block in a snapshot file holds a copy of the physical block data at the time of snapshot creation. A 'hole' in the snapshot file signifies that the snapshot's version of this block is identical to a later snapshot's version or to the version on the file system. Snapshot files are marked with a snapshot file flag, which is inherited from their parent directory and cannot be otherwise set on regular files.

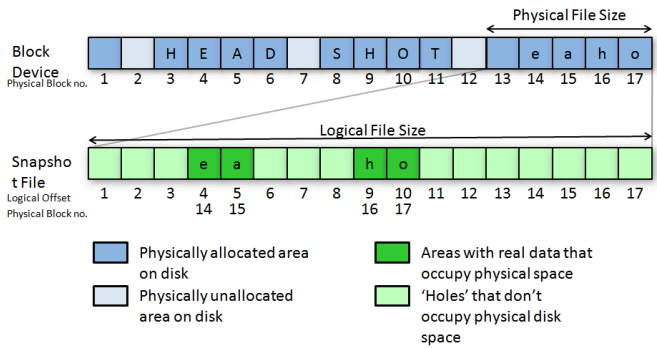

Fig 2.1 Sparse File

**Copy on Write (COW)**

Copy on write is a strategy in which a copy of data to be modified is made before write. In this method data from the original physical block is copied to a new physical location. The old data in the original physical block is then overwritten by new data. In case of a rewrite, after a snapshot is taken, the original data is preserved under snapshot file by the COW operation. Thus the logical offset in the snapshot file corresponding to the original physical block offset is mapped to the new physical location containing original data.

**Move on Write (MOW)**

Move on write is a slight variation of copy on write technique. In this case too copy of the data is made to a new physical location however, the new data is written at new physical location instead of the original physical location. Thus in case of rewrite after snapshot, logical offset in the snapshot file corresponds to the original physical location itself.

**COW Bitmap**

Cow bitmap keeps track of the blocks on which the COW or MOW operation is to be performed, in the same fashion as block bitmap keep track of the blocks which used by the file system. A COW bitmap is essentially a copy of the block bitmap at the time of creation of the snapshot and thus keeps track of the blocks which were present in the file system at that time. A 'set' bit in COW bitmap indicates that the corresponding block is in use by a snapshot.

**Exclude Bitmap**

Exclude Bitmap keeps track of the blocks which are to be skipped from any COW or MOW operation, essentially the blocks mapped to the snapshot file. The result is that snapshot file blocks are never copied or moved to active snapshot. During initialization of a COW bitmap block, the block bitmap block is masked with the exclude bitmap block.

**Extent**

Ext4 is an extent based file system, wherein extent mapped files can be used instead of the traditional block mapped files. An extent is a combination of two integers, the first stating the offset block and the second denoting the length i.e. number of contiguous blocks after the offset that are free or allocated. It thus improves large file performance and reduces fragmentation. A single extent in ext4 can map up to 128 MB of contiguous space with a 4 KB block size [2]. There can be 4 extents stored in the inode. When there are more than 4 extents to a file, the rest of the extents are indexed in an Htree.

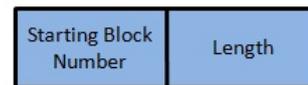

Fig.2.2 Extent

### 3. WORKING

Let us illustrate the working by the following example:

1.  Suppose initially the file system consists of a file which contains data "HEADSHOT". As shown in figure 3.1 the file consists of two data extents (40-43) "HEAD" and (50-53) "SHOT" referred to by its inode at block number 10.

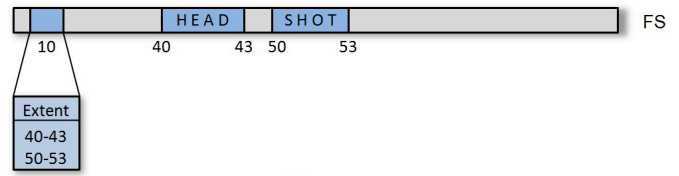

Fig. 3.1 File System

2.  Now at this stage if we take a snapshot S1 then, a new sparse file for the snapshot is created with its inode at block number 90 as shown in figure 3.2. This snapshot is the 'active' snapshot as it is the only one which is being modified. Initially the snapshot file is empty and consists of holes for corresponding data blocks. Thus the logical offsets (L) 10, 40-43, 50-53 map to physical offsets (P) '0' which represents a 'hole'.

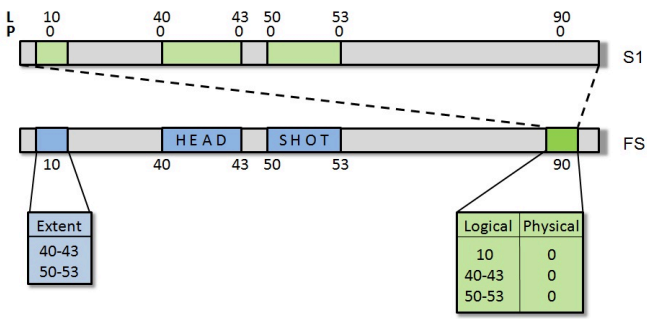

Fig. 3.2 Snapshot (S1) Take

3.     Suppose we have a re-write request for data at extent 40-43 from "HEAD" to "SNAP". Thus now we do a MOW operation for the data and COW operation for metadata (inode). As explained earlier, the data "SNAP" is written at a new location 60-63(MOW) and a copy of the inode at location 10 is made to the location 20 (COW). The metadata at location 10 is then modified to point extents "SNAP"(60-63) and "SHOT" (50-53). The logical offsets at 10 and 40-43 in the snapshot file are then mapped to blocks 20 and 40-43 instead of 'holes'. Thus the file refers to data "SNAPSHOT" under the file system and the same file refers to data "HEADSHOT" under snapshot S1.

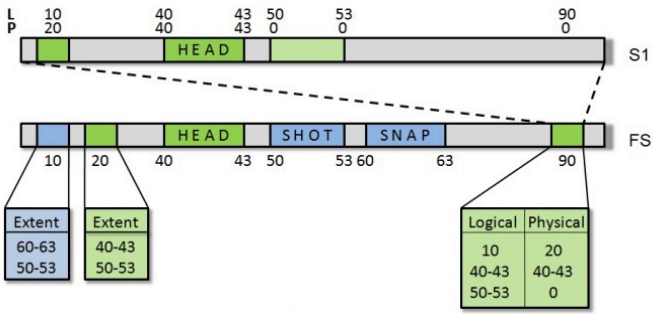

Fig. 3.3 Data Rewrite

4.     Now we take another snapshot S2, thus an empty snapshot file consisting of all holes mapping to data blocks on disk will be created (shown in figure 3.4) Now S2is the 'active' snapshot as explained earlier.

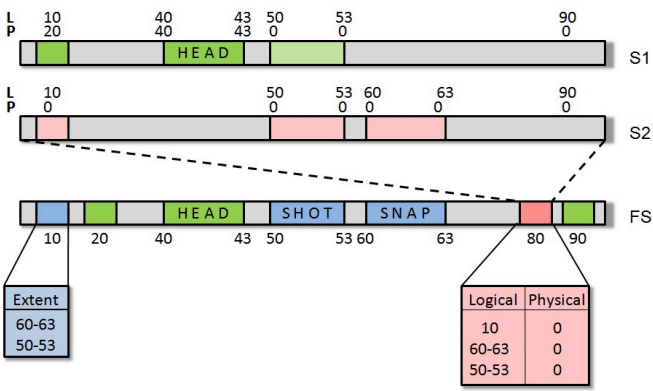

Fig. 3.4 Snapshot (S2) Take

5.     Suppose we have a re-write request for data from "SH" to "FS" at location 50-51. The data "FS" is written at a new location 70-71(MOW) and a copy of the inode at location 10 is made to the location 30(COW). The metadata at location 10 is then modified to point extents "SNAP"(60-63) "FS" (70-71) and "OT"(52-53). The logical offsets at 10 and 50-51 in the snapshot file S2 are then mapped to blocks 30 and 50-51 instead of 'holes'. Thus the file refers to data "SNAPFSOT" under the file system and the same file refers to data "SNAPSHOT" under snapshot S2.

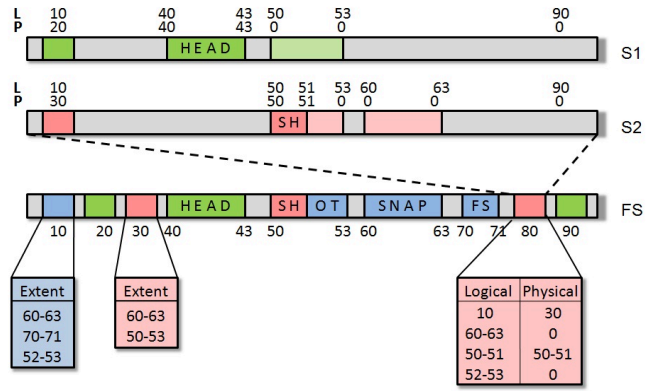

Fig. 3.5 Data Rewrite – Extent Break

6.     Suppose now there is a request to delete data "OT". Thus entry for extent 70-71in inode at 10 is removed and the inode points to only "SNAP" 60-63 and "FS" 50-51. The extent containing data "OT" is not freed but is protected under the active snapshot S2 by MOW operation. The snapshot file inode maps the logical offset52-53 to the blocks 52-53 instead of hole.

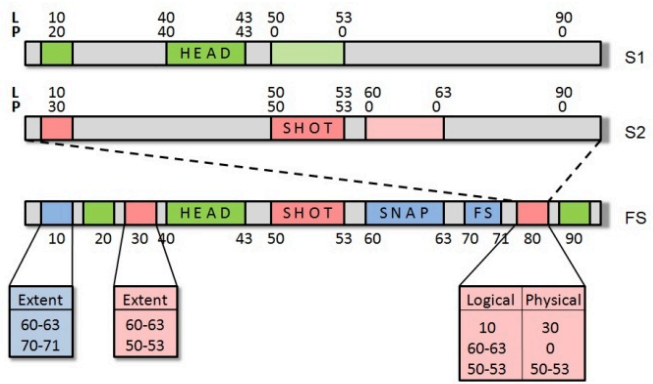

Fig. 3.6 Data Delete

7.     Now we wish to delete the snapshot S2, then the data blocks under S2 are checked for being referred to by any previous snapshots. In this case (as in figure 3.6) the data "SHOT"(50-53) is referred to by S1.The logical offsets 50-53 in the snapshot file S1 are then mapped to blocks 50-53 instead of 'holes'. The inode blocks at location 80 (S2) and 30 (File under S2) are then unallocated and the space is returned to the file system.(as shown in figure 3.7)

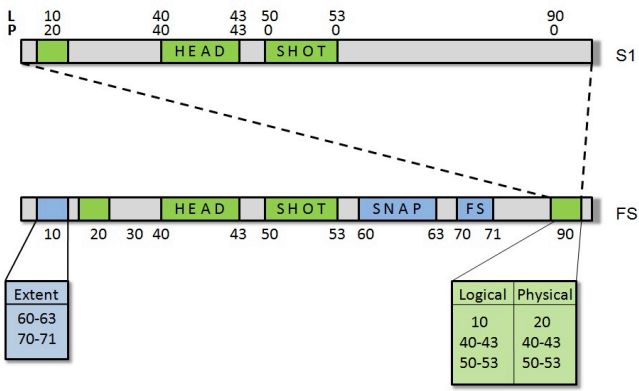

Fig. 3.7 Snapshot Delete

## 4. BENEFITS

- Snapshots are space efficient; require significantly less storage space as compared to traditional backup techniques.
- Creation of a snapshot (backup) is time efficient and requires low performance overhead.
- Some of the next generation file systems such as ZFS, Btrfs support snapshot feature, however both are not ready for production use and both implement COW by default. By implementing COW and MOW along with snapshot feature in Ext4, which is already a file system used in production system[5][6], Next4 implementation will add use case for Ext4.

## 5. CONCLUSION

In this paper, we present a file system level snapshot design suitable for recovery and backup. Snapshot technology represents one of the most significant storage enhancements in recent years, promising to reshape future data backup and recovery solutions. Apart from backup systems snapshot would also prove useful in applications such as software testing and system recovery tools. We implement snapshot in the extent based Ext4 file system. Ext4 is the successor of Ext3, the most widely used file system, and is shipped as the base file system in most Linux distributions today. Thus introducing snapshot feature in Ext4 will cater a huge number of users with its benefits.

## ACKNOWLEDGEMENT

We wish to thank the following people for their contribution to the system or the paper. Mrs. Rekha Kulkarni whose reviews helped us to build on our own ideas. Mr. Vedang Manerikar and Mr. Chinmay Kamat, our guides whose ideas and timely criticism helped us to bring this project about. Mr. Amir Golstein, who laid the seed of the idea in the first place. Lastly, but not the least, we wish to thank the anonymous reviewers, who patiently read our paper and gave us valuable inputs which helped make significant improvements in this paper.